\documentclass[english,pop,showpacs,reprint,superscriptaddress,nofootinbib]{revtex4-1}
\usepackage[T1]{fontenc}
\usepackage[latin9]{inputenc}
\usepackage{amsmath}
\usepackage{amssymb}
\usepackage{graphicx}

\makeatletter
\usepackage[colorlinks=true,citecolor=blue]{hyperref}

\makeatother

\usepackage{babel}
\begin{document}
\title{Nonlinear saturation and oscillations of collisionless zonal flows}
\author{Hongxuan Zhu}
\affiliation{Department of Astrophysical Sciences, Princeton University, Princeton,
NJ 08544 }
\affiliation{Princeton Plasma Physics Laboratory, Princeton, NJ 08543}
\author{Yao Zhou}
\affiliation{Princeton Plasma Physics Laboratory, Princeton, NJ 08543}
\author{I. Y. Dodin}
\affiliation{Department of Astrophysical Sciences, Princeton University, Princeton,
NJ 08544 }
\affiliation{Princeton Plasma Physics Laboratory, Princeton, NJ 08543}
\begin{abstract}
In homogeneous drift-wave (DW) turbulence, zonal flows (ZFs) can be
generated via a modulational instability (MI) that either saturates
monotonically or leads to oscillations of the ZF energy at the nonlinear
stage. This dynamics is often attributed as the predator\textendash prey
oscillations induced by ZF collisional damping; however, similar dynamics
is also observed in collisionless ZFs, in which case a different mechanism
must be involved. Here, we propose a semi-analytic theory that explains
the transition between the oscillations and saturation of collisionless
ZFs within the quasilinear Hasegawa\textendash Mima model. By analyzing
phase-space trajectories of DW quanta (driftons) within the geometrical-optics
(GO) approximation, we argue that the parameter that controls this
transition is $N\sim\gamma_{{\rm MI}}/\omega_{{\rm DW}}$, where $\gamma_{{\rm MI}}$
is the MI growth rate and $\omega_{{\rm DW}}$ is the linear DW frequency.
We argue that at $N\ll1$, ZFs oscillate due to the presence of so-called
passing drifton trajectories, and we derive an approximate formula
for the ZF amplitude as a function of time in this regime. We also
show that at $N\gtrsim1$, the passing trajectories vanish and ZFs
saturate monotonically, which can be attributed to phase mixing of
higher-order sidebands. A modification of $N$ that accounts for effects
beyond the GO limit is also proposed. These analytic results are tested
against both quasilinear and fully-nonlinear simulations. They also
explain the earlier numerical results by Connaughton\emph{ et al}.
{[}J. Fluid Mech. \textbf{654}, 207 (2010){]} and Gallagher \emph{et
al}. {[}Phys. Plasmas \textbf{19}, 122115 (2012){]} and offer a revised
perspective on what the control parameter is that determines the transition
from the oscillations to saturation of collisionless ZFs. 
\end{abstract}
\maketitle

\section{Introduction}

Zonal flows (ZFs) are banded sheared flows that can spontaneously
emerge from drift-wave (DW) turbulence in magnetized plasmas \citep{Lin98,Diamond05,Rogers00,Jenko00}
and, similarly, from Rossby-wave turbulence in the atmospheres of
rotating planets \citep{Vasavada05}. They are considered important
as regulators of turbulent transport and thus have been studied actively
by many researchers. One mechanism of the ZF generation is the secondary
(or zonostrophic) instability \citep{Rogers00,Smolyakov00a,Smolyakov00b,Strintzi07,Srinivasan2012,Parker2013,Parker2014,Parker14Thesis,St-Onge17a,Marston16},
and the modulational instability (MI) as a special case when the DW
is monochromatic \citep{Qi19,Chen00,Connaughton10,Gallagher12,Connaughton15,Champeaux01,Koshkarov16,Lashkin08,Jenko06,Shukla02,Manfredi01,Lashmore-Davies01,Gill74,Mahanti81,St-Onge17b}.
The linear stage of the MI is generally understood, but the dynamics
of ZFs at the nonlinear stage is not sufficiently explored. Some progress
in this area has been made by applying quasilinear (QL) models (Sec.~\ref{subsec:BasicEquation_QL}),
such as the second-order cumulant expansion theory (CE2) \citep{Srinivasan2012,Parker2013,Parker2014,Parker14Thesis,St-Onge17a,Marston16},
or the stochastic structural stability theory (SSST) \citep{Constantinou18,Farrell07,Bakas15};
however, those are not particularly intuitive. A more intuitive paradigm
was proposed based on a simpler QL model known as the wave-kinetic
equation (WKE) \citep{Smolyakov00a,Smolyakov00b,Ruiz16,Ruiz18,Zhu18a,Zhu18b,Parker16,Smolyakov99,Kaw2001}.
The WKE treats DW turbulence as a collection of DW quanta (``driftons''),
for which the ZF velocity serves as a collective field. Then, the
ZF\textendash DW interactions can be viewed as the predator\textendash prey
dynamics, which can result in either monotonic or oscillatory energy
exchange between the ZFs and DWs \citep{Malkov01,Miki12,Diamond94,Kim03,Berionni11}.
Such dynamics is indeed observed in simulations \citep{Kobayashi15}.
However, the existing paradigm substantially relies on collisional
damping of ZFs, whereas simulations indicate that nonlinear saturation
and oscillations are also possible when the ZF is collisionless \citep{Connaughton10,Gallagher12,Manfredi01,Lashmore-Davies01,Mahanti81}.
This leads to the question of how this dynamics can be explained within
a collisionless theory.

Here, we propose a semi-analytic theory that explains the transition
between the oscillations and saturation of collisionless ZFs within
the QL Hasegawa\textendash Mima model. By analyzing phase-space trajectories
of DW quanta (driftons) within the geometrical-optics (GO) approximation,
we argue that the parameter that controls this transition is $N\sim\gamma_{{\rm MI}}/\omega_{{\rm DW}}$,
where $\gamma_{{\rm MI}}$ is the MI growth rate and $\omega_{{\rm DW}}$
is the linear DW frequency. We argue that at $N\ll1$, ZFs oscillate
due to the presence of so-called passing drifton trajectories, and
we derive an approximate formula for the ZF amplitude as a function
of time in this regime. In doing so, we also extend the applicability
of the popular ``four-mode truncation'' (4MT) model \citep{Chen00,Connaughton10,Gallagher12,Connaughton15,Champeaux01,Koshkarov16,Lashkin08,Jenko06,Shukla02,Manfredi01,Lashmore-Davies01,Mahanti81,Gill74,St-Onge17b},
which is commonly used for the linear stage, to nonlinear ZF\textendash DW
interactions. We also show that at $N\gtrsim1$, the passing trajectories
vanish and ZFs saturate monotonically, which can be attributed to
phase mixing of higher-order sidebands. A modification of $N$ that
accounts for effects beyond the GO limit is also proposed. These analytic
results are tested against both QL and fully-nonlinear (NL) simulations.
They also explain the earlier numerical results by Connaughton \emph{et
al.} \citep{Connaughton10} and Gallagher \emph{et al.} \citep{Gallagher12}
and offer a revised perspective on what the control parameter is that
determines the transition from oscillations to saturation of collisionless
ZFs.

Our paper is organized as follows. In Sec.~\ref{sec:BasicEquation},
we introduce the basic equations, including the Hasegawa\textendash Mima
equation, the 4MT, the QL approximation, the Wigner function, and
the WKE that we use later on. In Sec.~\ref{sec:NonlinearMI}, we
discuss the nonlinear stage of the MI using both the 4MT description
and the WKE, and we also propose our control parameter $N$. In Sec.~\ref{sec:Simulations},
we compare our results with those in Refs~\citep{Connaughton10,Gallagher12}.
In Sec.~\ref{sec:conclusion} we summarize our main conclusions.

\section{Basic equations}

\label{sec:BasicEquation}

\subsection{Hasegawa\textendash Mima equation}

Both electrostatic DW turbulence in plasmas and Rossby-wave turbulence
in the atmospheres of rotating planets are often modeled by the Hasegawa\textendash Mima
equation (HME) \citep{Hasegawa77}:
\begin{gather}
\frac{\partial w}{\partial t}+(\hat{\boldsymbol{z}}\times\nabla\varphi)\cdot\nabla w+\beta\,\frac{\partial\varphi}{\partial x}=0,\label{eq:BasicEquation_HME}\\
w=(\nabla^{2}-L_{{\rm D}}^{-2}\hat{\alpha})\varphi.\label{eq:BasicEquation_vorticity}
\end{gather}
(In geophysics, it is also known as the Obukhov\textendash Charney
equation \citep{DolzhanskyBook}.) Here, the geophysics coordinate
convention is used to simplify comparisons with the earlier relevant
studies (Sec.~\ref{sec:Comparisons}). The HME describes wave turbulence
on a two-dimensional plane $(x,y)$, where ZFs develop along the $x$-direction,
and $\nabla^{2}\doteq\partial_{x}^{2}+\partial_{y}^{2}$ is the Laplacian.
In the plasma-physics context, the system is assumed to be immersed
in a uniform magnetic field along the $z$ axis, the plasma is assumed
to have a constant density gradient along the $y$ axis, $\beta$
is a scalar constant proportional to this gradient, $L_{{\rm D}}$
is the ion-sound radius, and $\varphi$ is the perturbation of the
electrostatic potential. In the geophysical context, the constant
$\beta$ is proportional to the latitudinal gradient of the vertical
rotation frequency, $L_{{\rm D}}$ is the deformation radius, and
$\varphi$ is the stream function.

The operator $\hat{\alpha}$ is an identity operator in the original
HME (oHME). The so-called modified HME (mHME), which is also known
as ``generalized'' \citep{Ruiz16,Krommes00} or ``extended'' HME
\citep{Connaughton15,Gallagher12}, uses $\hat{\alpha}f\doteq f-\langle f\rangle$,
where $f=f(x,y,t)$ is any field quantity and $\langle f\rangle\doteq\int_{0}^{L_{x}}fdx/L_{x}$
denotes the zonal average, where $L_{x}$ is the system length in
the $x$-direction. (For more details, see, for example, Ref.~\citep{St-Onge17b,Hammett93}.)
Below, we consider both the oHME and mHME and treat them on the same
footing.

\subsection{Fourier decomposition and the 4MT}

\label{subsec:BasicEquation_4MT}

It is common to approach the dynamics of $\varphi$ in the Fourier
representation, $\varphi=\sum_{\boldsymbol{k}}\varphi_{\boldsymbol{k}}(t)\exp(i\boldsymbol{k}\cdot\boldsymbol{x})$,
where $\boldsymbol{k}=(k_{x},k_{y})$ and $\boldsymbol{x}=(x,y)$.
This leads to the following equation for $\varphi_{\boldsymbol{k}}$:
\begin{gather}
\frac{\partial\varphi_{\boldsymbol{k}}}{\partial t}=i\omega_{\boldsymbol{k}}\varphi_{\boldsymbol{k}}+\frac{1}{2}\sum_{\boldsymbol{k}_{1},\boldsymbol{k}_{2}}T(\boldsymbol{k},\boldsymbol{k}_{1},\boldsymbol{k}_{2})\varphi_{\boldsymbol{k}_{1}}\varphi_{\boldsymbol{k}_{2}}\delta_{\boldsymbol{k},\boldsymbol{k}_{1}+\boldsymbol{k}_{2}}.\label{eq:BasicEquation_4MT_Fourier}
\end{gather}
Here, $\delta_{\boldsymbol{k},\boldsymbol{k}_{1}+\boldsymbol{k}_{2}}$
equals one only if $\boldsymbol{k}=\boldsymbol{k}_{1}+\boldsymbol{k}_{2}$
and is zero otherwise,
\begin{equation}
\omega_{\boldsymbol{k}}\doteq-\frac{\beta k_{x}}{k_{{\rm D}}^{2}}
\end{equation}
is the linear DW frequency, and 
\begin{equation}
T(\boldsymbol{k},\boldsymbol{k}_{1},\boldsymbol{k}_{2})\doteq-\frac{k_{1,{\rm D}}^{2}-k_{2,{\rm D}}^{2}}{k_{{\rm D}}^{2}}\,(\boldsymbol{k}_{1}\times\boldsymbol{k}_{2})\cdot\hat{\boldsymbol{z}}\label{eq:BasicEquation_4MT_coefficient}
\end{equation}
are the coefficients that govern the nonlinear mode coupling. Also,
\begin{equation}
k_{{\rm D}}^{2}\doteq|\boldsymbol{k}|^{2}+L_{{\rm D}}^{-2}\alpha_{\boldsymbol{k}},
\end{equation}
and $k_{n,{\rm D}}^{2}$ (where $n=1,2$) are similarly defined as
\begin{equation}
k_{n,{\rm D}}^{2}\doteq|\boldsymbol{k}_{n}|^{2}+L_{{\rm D}}^{-2}\alpha_{\boldsymbol{k}_{n}}.\label{eq:BasicEquation_4MT_knD}
\end{equation}
And finally, $\alpha_{\boldsymbol{k}}$ is the Fourier representation
of the operator $\hat{\alpha}$; namely, for the oHME $\alpha_{\boldsymbol{k}}$
is unity, and for the mHME $\alpha_{\boldsymbol{k}}$ equals zero
if $k_{x}=0$ and equals unity if $k_{x}\neq0$.

It is also common to introduce the 4MT, which is a truncation of the
system (\ref{eq:BasicEquation_4MT_Fourier}) that retains only four
Fourier harmonics, namely, those with wave vectors $\boldsymbol{p}$,
$\boldsymbol{q}$, and $\boldsymbol{p}_{\pm}\doteq\boldsymbol{p}\pm\boldsymbol{q}$.
(Since $\varphi$ is real, one has $\varphi_{-\boldsymbol{k}}=\varphi_{\boldsymbol{k}}^{*}$;
hence, the harmonics with wave vectors $-\boldsymbol{p}$, $-\boldsymbol{q}$,
and $-\boldsymbol{p}_{\pm}$ are included too.) Then, for $\Phi_{\boldsymbol{k}}\doteq\varphi_{\boldsymbol{k}}\exp(-i\omega_{\boldsymbol{k}}t)$,
Eq.~(\ref{eq:BasicEquation_4MT_Fourier}) gives 
\begin{subequations}
\begin{gather}
\partial_{t}\Phi_{\boldsymbol{p}}=T(\boldsymbol{p},\boldsymbol{q},\boldsymbol{p}_{-})\Phi_{\boldsymbol{q}}\Phi_{\boldsymbol{p}_{-}}e^{i\Delta_{-}t}\nonumber \\
+T(\boldsymbol{p},-\boldsymbol{q},\boldsymbol{p}_{+})\Phi_{\boldsymbol{q}}^{*}\Phi_{\boldsymbol{p}_{+}}e^{i\Delta_{+}t},\label{eq:BasicEquation_4MT1}\\
\partial_{t}\Phi_{\boldsymbol{q}}=T(\boldsymbol{q},\boldsymbol{p},-\boldsymbol{p}_{-})\Phi_{\boldsymbol{p}}\Phi_{\boldsymbol{p}_{-}}^{*}e^{-i\Delta_{-}t}\nonumber \\
+T(\boldsymbol{q},-\boldsymbol{p},\boldsymbol{p}_{+})\Phi_{\boldsymbol{p}}^{*}\Phi_{\boldsymbol{p}_{+}}e^{i\Delta_{+}t},\label{eq:BasicEquation_4MT2}\\
\partial_{t}\Phi_{\boldsymbol{p}_{-}}=T(\boldsymbol{p}_{-},\boldsymbol{p},-\boldsymbol{q})\Phi_{\boldsymbol{p}}\Phi_{\boldsymbol{q}}^{*}e^{-i\Delta_{-}t},\label{eq:BasicEquation_4MT3}\\
\partial_{t}\Phi_{\boldsymbol{p}_{+}}=T(\boldsymbol{p}_{+},\boldsymbol{p},\boldsymbol{q})\Phi_{\boldsymbol{p}}\Phi_{\boldsymbol{q}}e^{-i\Delta_{+}t},\label{eq:BasicEquation_4MT4}
\end{gather}
\label{eq:BasicEquation_4MTeq}
\end{subequations}
 where $\Delta_{\pm}=\omega_{\boldsymbol{p}}\pm\omega_{\boldsymbol{q}}-\omega_{\boldsymbol{p}_{\pm}}$.

\subsection{Modulational instability}

Suppose a perturbation on a primary wave $\Phi_{\boldsymbol{p}}=\Phi_{0}$,
\begin{gather}
\left(\begin{array}{c}
\Phi_{\boldsymbol{p}}\\
\Phi_{\boldsymbol{q}}\\
\Phi_{\boldsymbol{p}_{+}}\\
\Phi_{\boldsymbol{p}_{-}}
\end{array}\right)=\left(\begin{array}{c}
\Phi_{0}\\
0\\
0\\
0
\end{array}\right)+\varepsilon\left(\begin{array}{c}
0\\
A_{\boldsymbol{q}}e^{-i\Omega_{\boldsymbol{q}}t}\\
A_{\boldsymbol{p}_{+}}e^{-i\Omega_{\boldsymbol{p}_{+}}t}\\
A_{\boldsymbol{p}_{-}}e^{-i\Omega_{\boldsymbol{p}_{-}}t}
\end{array}\right),
\end{gather}
where $\varepsilon$ is small. Then, the linearized Eq.~(\ref{eq:BasicEquation_4MTeq})
gives $\Omega_{\boldsymbol{p}\pm}=\Delta_{\pm}\pm\Omega_{\boldsymbol{q}}$
together with the following dispersion relation:
\begin{multline}
\Omega_{\boldsymbol{q}}+\frac{|\Phi_{0}|^{2}T(\boldsymbol{q},-\boldsymbol{p},\boldsymbol{p}_{+})T(\boldsymbol{p}_{+},\boldsymbol{q},\boldsymbol{p})}{\Delta_{+}+\Omega_{\boldsymbol{q}}}\\
-\frac{|\Phi_{0}|^{2}T(\boldsymbol{q},-\boldsymbol{p}_{-},\boldsymbol{p})T(\boldsymbol{p}_{-},-\boldsymbol{q},\boldsymbol{p})}{\Delta_{-}-\Omega_{\boldsymbol{q}}}=0.
\end{multline}
(The derivation can be found, for example, in Ref.~\citep{Gallagher12}.)
This equation can have a complex solution for $\Omega_{\boldsymbol{q}}$
with $\text{Im}\,\Omega_{\boldsymbol{q}}>0$, which signifies the
presence of the MI. As shown in Refs.~\citep{Connaughton10,Gill74},
the 4MT is indeed often a good approximation at the linear stage of
the MI.

In the following, we restrict our discussions to the case when $\boldsymbol{p}=(p,0)$
and $\boldsymbol{q}=(0,q)$. In this case, $\omega_{\boldsymbol{q}}=0$
and
\begin{equation}
\Delta_{\pm}\equiv\Delta\doteq\frac{\beta pq^{2}}{p_{{\rm D}}^{2}(p_{{\rm D}}^{2}+q^{2})},\label{eq:BasicEquation_Delta}
\end{equation}
where
\begin{equation}
p_{{\rm D}}^{2}\doteq L_{{\rm D}}^{-2}+p^{2}.
\end{equation}
Then, one finds that $\Omega_{\boldsymbol{q}}^{2}$ is real, and hence
the MI has a positive growth rate $\gamma_{{\rm MI}}$ if $\Omega_{\boldsymbol{q}}^{2}$
is negative; namely, $\gamma_{{\rm MI}}^{2}\doteq-\Omega_{\boldsymbol{q}}^{2}$,
which can be explicitly written as 
\begin{equation}
\gamma_{{\rm MI}}^{2}=\frac{2|\Phi_{0}|^{2}p^{2}q^{2}}{(1+\alpha L_{{\rm D}}^{-2}q^{-2})}\,\frac{\delta'-1}{\delta+1}-\Delta^{2}.\label{eq:BasicEquation_gamma}
\end{equation}
Here, 
\begin{equation}
\delta\doteq\frac{p_{{\rm D}}^{2}}{q^{2}},\quad\delta'\doteq\frac{p_{{\rm D}}^{2}-\alpha L_{{\rm D}}^{-2}}{q^{2}},\label{eq:BasicEquation_delta_deltap}
\end{equation}
with $\alpha=1$ for the oHME and $\alpha=0$ for the mHME. Notably,
this implies that the mHME gives much larger growth rates than the
oHME does. Also, a necessary condition for the modulation to be unstable
is $\delta'>1$, which requires
\begin{equation}
q^{2}<p_{{\rm D}}^{2}-\alpha L_{{\rm D}}^{-2}.
\end{equation}

\subsection{Quasilinear approximation}

\label{subsec:BasicEquation_QL}

In order to describe the nonlinear stage of the MI analytically, we
proceed as follows. Let us decompose the field quantities into the
zonal-averaged part and the fluctuation part, $f=\langle f\rangle+\tilde{f}$.
Then, Eqs.~(\ref{eq:BasicEquation_HME}) and (\ref{eq:BasicEquation_vorticity})
become \citep{Parker14Thesis}
\begin{gather}
\frac{\partial\tilde{w}}{\partial t}+U\,\frac{\partial\tilde{w}}{\partial x}+\left[\beta-\left(\frac{\partial^{2}}{\partial y^{2}}-\alpha L_{{\rm D}}^{-2}\right)U\right]\frac{\partial\tilde{\varphi}}{\partial x}=f_{{\rm NL}},\label{eq:BasicEquation_DW}\\
\left[1-\alpha L_{{\rm D}}^{-2}\left(\frac{\partial}{\partial y}\right)^{-2}\right]\frac{\partial U}{\partial t}=-\frac{\partial}{\partial y}\langle\tilde{v}_{x}\tilde{v}_{y}\rangle.\label{eq:BasicEquation_ZF}
\end{gather}
Here, $U(y,t)\doteq-\partial_{y}\langle\varphi\rangle$ is the ZF
velocity, $\tilde{\boldsymbol{v}}\doteq\hat{\boldsymbol{z}}\times\nabla\tilde{\varphi}$
is the fluctuation velocity, $\partial_{y}^{-2}$ is an operator that
in the wave-vector (Fourier) representation is simply a multiplication
by $-k_{y}^{-2}$, and $f_{{\rm NL}}\doteq\tilde{\boldsymbol{v}}\cdot\nabla\tilde{w}-\langle\tilde{\boldsymbol{v}}\cdot\nabla\tilde{w}\rangle$
describes self-interactions of DWs, or eddy\textendash eddy interactions.
We shall simplify the problem by ignoring these interactions, i.e.,
by adopting $f_{{\rm NL}}=0$. This is the commonly-used QL approximation,
which often yields an adequate description of ZF\textendash DW interactions
\citep{Parker14Thesis}. (We shall also discuss the applicability
of this approximation in Sec.~\ref{subsec:Simulations_NL}). Notably,
once the QL approximation is adopted, the 4MT model becomes the exact
description of the linear MI.

We also assume, for simplicity, that the ZF is sinusoidal and non-propagating
(assuming $\gamma_{{\rm MI}}$ is real),
\begin{equation}
U=u(t)\cos qy,\label{eq:BasicEquation_cosU}
\end{equation}
where for clarity we choose the origin on the $y$ axis such that
$y=0$ corresponds to the maximum of $U$. (Propagating zonal structures
are studied in detail in our Ref.~\citep{soliton}.) The assumption
of spatially-monochromatic ZF holds approximately if the ZF's second
and higher harmonics do not outpace the fundamental harmonic during
the linear MI. Hence, the ansatz (\ref{eq:BasicEquation_cosU}) implies
that the value of $q$ is close to the one that maximizes $\gamma_{{\rm MI}}$.
Then, the DW field can be represented as $\tilde{\varphi}={\rm Re}\sum_{m=-\infty}^{\infty}\varphi_{m}(t)\exp(ipx+imqy)$,
where $\varphi_{0}(t=0)=\Phi_{0}$ is the primary-wave amplitude,
and the DW equation (\ref{eq:BasicEquation_DW}) becomes
\begin{multline}
\frac{\partial\varphi_{m}}{\partial t}=i\omega_{m}\varphi_{m}+\frac{ipu}{2}\left(\frac{q^{2}+\alpha L_{{\rm D}}^{-2}-k_{m+1,{\rm D}}^{2}}{k_{m,{\rm D}}^{2}}\right)\varphi_{m+1}\\
+\frac{ipu}{2}\left(\frac{q^{2}+\alpha L_{{\rm D}}^{-2}-k_{m-1,{\rm D}}^{2}}{k_{m,{\rm D}}^{2}}\right)\varphi_{m-1},\label{eq:BasicEquation_Fourier_DW}
\end{multline}
where 
\begin{equation}
\omega_{m}\doteq-\frac{\beta p}{k_{m,{\rm D}}^{2}},\quad k_{m,{\rm D}}^{2}\doteq p_{{\rm D}}^{2}+(mq)^{2}.
\end{equation}
Note that the definition of $k_{m,D}$ is consistent with that given
by Eq.~(\ref{eq:BasicEquation_4MT_knD}).

\subsection{Wigner function and WKE}

Equation (\ref{eq:BasicEquation_Fourier_DW}) describes the mode-coupling
among different Fourier modes of DWs due to the ZF. Now, we seek to
interpret this equation in terms of the drifton phase-space dynamics.
Consider the zonal-averaged Wigner function of DWs,
\begin{multline}
W(y,\boldsymbol{k},t)=\int d^{2}se^{-i\boldsymbol{k}\cdot\boldsymbol{s}}\left<\tilde{w}(\boldsymbol{x}+\frac{\boldsymbol{s}}{2},t)\tilde{w}(\boldsymbol{x}-\frac{\boldsymbol{s}}{2},t)\right>.\label{eq:BasicEquation_Wigner}
\end{multline}
It can be understood as the spectral representation of the DW two-point
correlation function, such as the one used in the CE2. It can also
be viewed as the quasiprobability of the drifton distribution in the
$(y,\boldsymbol{k})$ space, where the prefix ``quasi'' reflects
the fact that $W$ is not necessarily positive-definite. That said,
it becomes positive-definite in the GO limit defined as (i) $\alpha L_{{\rm D}}^{-2}+q^{2}\ll L_{{\rm D}}^{-2}+p^{2}$
and (ii) $\partial_{k_{y}}W\ll q^{-1}W$, when it can be considered
as the true distribution function of driftons \citep{Cartwright76}.
Then, the Wigner function can be shown to satisfy the following partial
differential equation \citep{Ruiz16,Parker16}:
\begin{equation}
\partial_{t}W(y,\boldsymbol{k},t)=\{\mathcal{H},W\}+2\Gamma W.\label{eq:BasicEquation_WKE}
\end{equation}
Here, $\{\cdot,\cdot\}$ is the Poisson bracket,
\begin{equation}
\{A,B\}\doteq\frac{\partial A}{\partial\boldsymbol{x}}\cdot\frac{\partial B}{\partial\boldsymbol{k}}-\frac{\partial A}{\partial\boldsymbol{k}}\cdot\frac{\partial B}{\partial\boldsymbol{x}},
\end{equation}
and the functions $\mathcal{H}$ and $\Gamma$ can be interpreted
as the drifton Hamiltonian and the drifton dissipation rate, respectively.
Specifically, they are given by \citep{Ruiz16,Zhu18b}
\begin{gather}
\mathcal{H}(y,\boldsymbol{k},t)=k_{x}\left[u(t)\cos qy-\frac{\beta+(q^{2}+\alpha L_{{\rm D}}^{-2})u(t)\cos qy}{k_{{\rm D}}^{2}}\right],\nonumber \\
\Gamma(y,\boldsymbol{k},t)=-\frac{k_{x}k_{y}q(q^{2}+\alpha L_{{\rm D}}^{-2})u(t)\sin qy}{k_{{\rm D}}^{4}}.\label{eq:BasicEquation_Wigner_hamiltonian}
\end{gather}
Equation (\ref{eq:BasicEquation_WKE}) is the WKE that describes the
phase-space dynamics of driftons (DW quanta). They obey Hamilton's
equations and move on constant-$\mathcal{H}$ surfaces in the phase-space
$(y,k_{y})$ if the ZF is stationary. (The $x$-momentum $k_{x}=p$
serves only as a parameter.) In the GO limit, $\Gamma$ is small and
unimportant for the discussions below. However, in general, keeping
$\Gamma$ is necessary to ensure that the WKE preserves the conservation
of the fundamental integrals of the HME \citep{Parker16}.

\subsection{Evolution of $\boldsymbol{W}$ beyond the GO approximation}

Beyond the GO approximation, the Wigner function satisfies a pseudo-differential
equation known as the Wigner\textendash Moyal equation \citep{Ruiz16}.
But here, we use a somewhat different formulation. Following the procedure
in Ref.~\citep{Ruiz16}, we consider the \emph{spectrum} of $W$,
\begin{equation}
W_{\lambda}(\boldsymbol{k},t)\doteq\int W(y,\boldsymbol{k},t)\exp(-i\lambda y)dy,\label{eq:BasicEquation_Wigner_Spectral}
\end{equation}
which is a function in the double-momentum space $(\lambda,k_{y})$.
Then, Eq.~(\ref{eq:BasicEquation_Fourier_DW}) is transformed into
\begin{multline}
\partial_{t}W_{\lambda}(p,k_{y},t)=i\beta p\left(\frac{1}{\kappa_{+\lambda}^{2}}-\frac{1}{\kappa_{-\lambda}^{2}}\right)W_{\lambda,0}\\
-\frac{iup}{2}\,Q_{\lambda-2q}W_{\lambda-q,-q}+\frac{iup}{2}\,Q_{2q-\lambda}W_{\lambda-q,+q}\\
-\frac{iup}{2}\,Q_{\lambda+2q}W_{\lambda+q,+q}+\frac{iup}{2}\,Q_{-2q-\lambda}W_{\lambda+q,-q},\label{eq:BasicEquation_Wigner_fullwave}
\end{multline}
where we introduced
\begin{gather*}
W_{a,b}\doteq W_{a}\left(p,k_{y}+\frac{b}{2},t\right),\quad\kappa_{a}^{2}\doteq p_{{\rm D}}^{2}+\left(k_{y}+\frac{a}{2}\right)^{2},
\end{gather*}
and
\begin{equation}
Q_{a}\doteq1-\frac{\alpha L_{{\rm D}}^{-2}+q^{2}}{\kappa_{a}^{2}}.\label{eq:BasicEquation_Wigner_Q}
\end{equation}

Equation (\ref{eq:BasicEquation_Wigner_fullwave}) is equivalent to
Eq.~(\ref{eq:BasicEquation_Fourier_DW}), but describes the dynamics
in the double-momentum space. The DW momentum flux can also be expressed
through $W_{\lambda}$, whose gradient drives the Fourier component
of the ZF, $U_{\lambda=q}\doteq\pi u(t)$. Specifically, from Eq.~(\ref{eq:BasicEquation_ZF}),
the following equation is obtained:
\begin{equation}
\frac{\partial U_{q}}{\partial t}=\frac{i}{(1+\alpha L_{{\rm D}}^{-2}q^{-2})}\int\frac{d^{2}k}{(2\pi)^{2}}\frac{k_{x}k_{y}q}{\kappa_{+q}^{2}\kappa_{-q}^{2}}W_{q}(\boldsymbol{k},t).\label{eq:BasicEquation_Uq}
\end{equation}

\section{Nonlinear stage of the modulation instability}

\label{sec:NonlinearMI}

\subsection{4MT description}

In order to describe the nonlinear stage of the MI, let us first consider
this instability within the 4MT model. We shall show that the 4MT
has an \emph{exact analytic solution} that not only gives the linear
growth rate (\ref{eq:BasicEquation_gamma}), but also predicts the
reversal of the ZF growth. We shall also propose a toy-model modification
of the 4MT that qualitatively explains the transition from ZF oscillations
to saturation.

For the MI, the initial condition is $\tilde{\varphi}(\boldsymbol{x},t=0)=\Phi_{0}\exp(ipx)+{\rm c.c.}$,
which corresponds to a delta function in the double-momentum space
{[}Eq.~(\ref{eq:BasicEquation_Wigner}){]},
\begin{gather}
W_{\lambda}(\boldsymbol{k},t=0)=a_{0}\delta(\lambda)\delta(k_{y})[\delta(k_{x}-p)+\delta(k_{x}+p)],\label{eq:NonlinearMI_W(t=00003D0)}\\
a_{0}\doteq8\pi^{3}(L_{{\rm D}}^{-2}+p^{2})^{2}|\Phi_{0}|^{2}.\label{eq:NonlinearMI_a0}
\end{gather}
According to Eq.~(\ref{eq:BasicEquation_Wigner_fullwave}), $W_{\lambda}(\boldsymbol{k},t)$
remains delta-shaped also at $t>0$, so we search it in the form
\begin{equation}
W_{\lambda}(p,k_{y},t)=\sum_{m,n}\bar{W}_{m,n}(t)\delta(\lambda-mq)\delta(k_{y}-nq/2),\label{eq:NonlinearMI_4MT_deltas}
\end{equation}
where the factor $\delta(k_{x}-p)$ is omitted, and $W_{\lambda}(-\boldsymbol{k},t)=W_{\lambda}(\boldsymbol{k},t)$.
The 4MT model (Sec.~\ref{subsec:BasicEquation_4MT}) corresponds
to keeping the following nine peaks:
\begin{gather*}
\bar{W}_{0,0}\doteq a(t),\quad\bar{W}_{\pm2,0}=\bar{W}_{0,\pm2}\doteq d(t),\\
\bar{W}_{1,\pm1}\doteq b(t)\pm ic(t),\quad\bar{W}_{-1,\pm1}\doteq b(t)\mp ic(t).
\end{gather*}
Here, $a$, $b$, $c$, and $d$ are real, and the initial value of
$a(t)$ is $a_{0}$. Using Eq.~(\ref{eq:BasicEquation_Wigner_fullwave}),
one obtains
\begin{gather}
\dot{a}=2puc\,\frac{\delta'}{1+\delta},\quad\dot{b}=\Delta c,\quad\dot{d}=puc\,\frac{1-\delta'}{\delta},\nonumber \\
\dot{c}=-\Delta b+\frac{pua}{2}\,\frac{1-\delta'}{\delta}+pud\,\frac{\delta'}{1+\delta}.\label{eq:NonlinearMI_4MT_abcd}
\end{gather}
Here, $\Delta$, $\delta$, and $\delta'$ are defined by Eqs.~(\ref{eq:BasicEquation_Delta})
and (\ref{eq:BasicEquation_delta_deltap}). Also, the ZF amplitude
is governed by {[}Eq.~(\ref{eq:BasicEquation_Uq}){]}
\begin{equation}
\dot{u}=-\frac{pq^{2}c}{\epsilon p_{{\rm D}}^{2}(p_{{\rm D}}^{2}+q^{2})},\label{eq:NonlinearMI_4MT_u}
\end{equation}
where we introduced
\begin{equation}
\epsilon\doteq2\pi^{3}(1+\alpha L_{{\rm D}}^{-2}q^{-2}).
\end{equation}
Note that DWs with $k_{x}=p$ and $k_{x}=-p$ contribute equally to
$\dot{u}$. 

In the following, we simplify the above equations and obtain the time-evolution
of $u$. First, notice that 
\begin{equation}
\frac{\dot{d}}{\dot{a}}=\frac{(1-\delta')(1+\delta)}{2\delta\delta'}.
\end{equation}
Thus, $d$ can be expressed through $a$:
\begin{equation}
d=\frac{(1-\delta')(1+\delta)}{2\delta\delta'}(a-a_{0}),
\end{equation}
where $a_{0}$ is given by Eq.~(\ref{eq:NonlinearMI_a0}). Next,
we introduce dimensionless variables
\begin{gather*}
\tau=t/T,\quad\bar{a}=a/A,\quad\bar{b}=b/B,\quad\bar{c}=c/C,\quad\bar{u}=u/V,
\end{gather*}
where we choose $B=C=\beta\epsilon V$ and 
\begin{gather*}
T=\frac{q^{2}\delta(\delta+1)}{\beta p},\,A=\frac{\beta^{2}\epsilon}{q^{2}(\delta+1)(\delta'-1)},\,V=\sqrt{\frac{A}{2\epsilon q^{2}\delta\delta'}}.
\end{gather*}
Then, Eqs.~(\ref{eq:NonlinearMI_4MT_abcd}) and (\ref{eq:NonlinearMI_4MT_u})
become
\begin{gather}
\frac{d\bar{a}}{d\tau}=\bar{u}\bar{c},\quad\frac{d\bar{u}}{d\tau}=-\bar{c},\nonumber \\
\frac{d\bar{b}}{d\tau}=\bar{c},\quad\frac{d\bar{c}}{d\tau}=-\bar{b}-\left(\bar{a}-\frac{\bar{a}_{0}}{2}\right)\bar{u},\label{eq:NonlinearMI_4MT_rescale}
\end{gather}
where $\bar{a}_{0}=a_{0}/A$. This leads to 
\begin{multline*}
\frac{d^{3}\bar{u}}{d\tau^{3}}=-\frac{d^{2}\bar{c}}{d\tau^{2}}=\frac{d\bar{b}}{d\tau}+\left(\bar{a}-\frac{\bar{a}_{0}}{2}\right)\frac{d\bar{u}}{d\tau}+\frac{d\bar{a}}{d\tau}u\\
=-\left[1-\left(\bar{a}-\frac{\bar{a}_{0}}{2}\right)\right]\frac{d\bar{u}}{d\tau}-\left(-\frac{d\bar{a}}{d\tau}u\right)=-\frac{d}{d\tau}(\bar{f}\bar{u}),
\end{multline*}
where 
\begin{equation}
\bar{f}\doteq1-\left(\bar{a}-\frac{\bar{a}_{0}}{2}\right).
\end{equation}
Assuming infinitesimally small initial $\bar{u}$, we have
\begin{equation}
\frac{d^{2}\bar{u}}{d\tau^{2}}+\bar{f}\bar{u}=0.
\end{equation}
Finally, since
\begin{equation}
\frac{d\bar{f}}{d\tau}=-\frac{d\bar{a}}{d\tau}=-\bar{u}\bar{c}=\frac{1}{2}\frac{d(\bar{u}^{2})}{d\tau},
\end{equation}
one has that $\bar{f}-\bar{u}^{2}/2$ is conserved; namely,
\begin{equation}
\bar{f}-\frac{\bar{u}^{2}}{2}=-g^{2}\doteq1-\frac{\bar{a}_{0}}{2}
\end{equation}
is a constant. (Having $g^{2}<0$ is possible but does not lead to
an instability; see below.) Hence, we obtain that $\bar{u}$ satisfies
a nonlinear-oscillator equation
\begin{equation}
\frac{d^{2}\bar{u}}{d\tau^{2}}=-\frac{d\Theta}{d\bar{u}},\quad\Theta(\bar{u})\doteq-\frac{g^{2}\bar{u}^{2}}{2}+\frac{\bar{u}^{4}}{8}.\label{eq:NonlinearMI_4MT_nonlinear}
\end{equation}

\begin{figure}
\includegraphics[width=1\columnwidth]{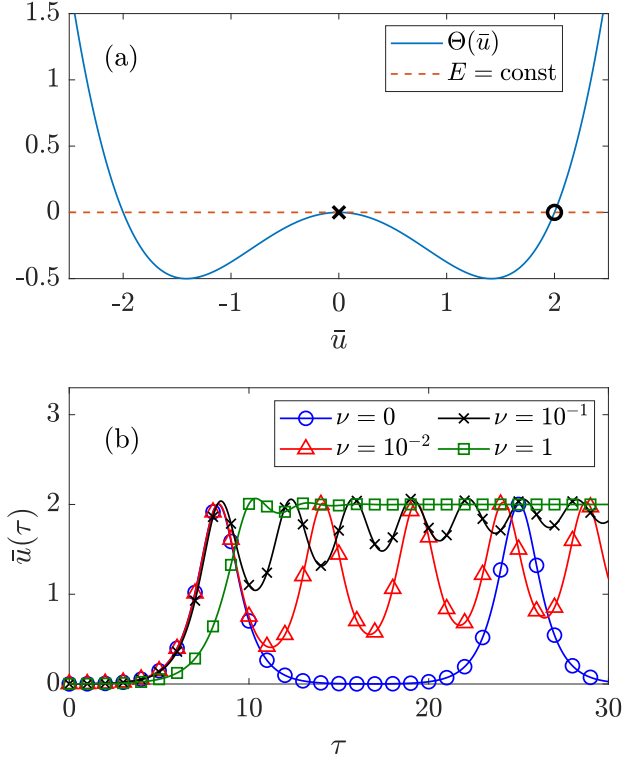}

\caption{(a) The effective potential $\Theta(\bar{u})$ {[}Eq.~(\ref{eq:NonlinearMI_4MT_nonlinear}){]}
with $\bar{a}_{0}=4$ (and hence $g^{2}=1>0$). The MI corresponds
to a initial condition $\bar{u}=\bar{u}_{0}$ near the origin (black
cross, $\bar{u}_{0}=10^{-3}$). At the nonlinear stage, $\bar{u}$
is constrained by the conservation of the ``energy'', $E=(d_{\tau}\bar{u})^{2}/2+\Theta$
(dashed line), and hence will start to decrease when reaching $\bar{u}\approx2$
(black circle). (b) Numerical solutions of the rescaled 4MT system
governed by Eqs.~(\ref{eq:NonlinearMI_4MT_rescale}) and (\ref{eq:NonlinearMI_4MT_damping}).
The coefficient $\nu$ in Eq.~(\ref{eq:NonlinearMI_4MT_damping})
describes an \emph{ad hoc} damping that mimics the coupling to other
DW sidebands beyond the 4MT. The initial conditions are $\bar{a}=\bar{a}_{0}=4$,
$\bar{u}=10^{-3}$, and $\bar{b}=\bar{c}=0$. A transition from oscillations
to saturation of $\bar{u}$ is observed as $\nu$ increases. \label{fig:4MT}}
\end{figure}

The effective potential $\Theta(\bar{u})$ is plotted in Fig.~\ref{fig:4MT}(a).
(A small nonzero initial value of $\bar{u}$ causes a slight alteration
of $\Theta$, but the qualitative picture remains the same.) The steady-state
solution $\bar{u}=0$ is unstable if $g^{2}>0$, which signifies the
presence of a linear instability (namely, the MI) with the growth
rate 
\begin{equation}
\gamma_{{\rm MI}}^{2}=\frac{g^{2}}{T^{2}}=\frac{1}{T^{2}}\left(\frac{a_{0}}{2A}-1\right),
\end{equation}
which is in agreement with Eq.~(\ref{eq:BasicEquation_gamma}). Beyond
the linear regime, i.e., when $\bar{u}^{4}$ is no longer negligible,
Eq.~(\ref{eq:NonlinearMI_4MT_nonlinear}) can also be integrated
exactly, yielding
\begin{equation}
\bar{u}=2g\,{\rm sech}(g\tau),\quad g=\sqrt{\frac{\bar{a}_{0}}{2}-1}.\label{eq:NonlinearMI_4MT_exact}
\end{equation}
This solution corresponds to the initial condition $\bar{a}(\tau\to-\infty)=\bar{a}_{0}$
and $\bar{u}(\tau\to-\infty)=0$, and the origin on the time axis
is chosen such that the ZF attains the maximum amplitude at $\tau=0$.
(A comparison between this analytic solution and simulations is presented
in Sec.~\ref{subsec:Simulations_Scan}) In our original variables,
this maximum amplitude is given by 
\begin{equation}
u_{{\rm max}}=2Vg=2VT\gamma_{{\rm MI}},
\end{equation}
or more explicitly,
\begin{equation}
u_{{\rm max}}=\sqrt{\frac{2\delta(\delta+1)}{\delta'(\delta'-1)}}\,\frac{\gamma_{{\rm MI}}}{p}.\label{eq:NonlinearMI_4MT_umax}
\end{equation}

The 4MT dynamics is numerically illustrated in Fig.~\ref{fig:4MT}(b).
Unlike the exact solution (\ref{eq:NonlinearMI_4MT_exact}), a finite
initial perturbation of $\bar{u}$ results in oscillations with a
finite period {[}Fig.~\ref{fig:4MT}(b), blue curve with circles{]}.
We also propose a toy model to illustrate the transition from oscillations
to saturation of $\bar{u}$, which adds an \emph{ad hoc} damping of
the sidebands to mimic their coupling to higher harmonics. Specifically,
we replace Eq.~(\ref{eq:NonlinearMI_4MT_rescale}) with 
\begin{equation}
\frac{d\bar{b}}{d\tau}=\bar{c}-\nu\bar{b},\quad\frac{d\bar{c}}{d\tau}=-\bar{b}-\left(\bar{a}-\frac{a_{0}}{2}\right)\bar{u}-\nu\bar{c},\label{eq:NonlinearMI_4MT_damping}
\end{equation}
where $\nu$ is some positive constant. Then, as we increase $\nu$,
a gradual transition from oscillations to saturation of $\bar{u}$
is observed as shown in Fig.~\ref{fig:4MT}(b). In particular, at
very large $\nu$ such that $\nu\gg d/d\tau$, one has
\begin{equation}
\bar{b}\approx\frac{\bar{c}}{\nu},\quad\bar{c}\approx-\frac{\bar{u}}{\nu}\left(\bar{a}-\frac{\bar{a}_{0}}{2}\right),
\end{equation}
and Eq.~(\ref{eq:NonlinearMI_4MT_rescale}) gives
\begin{equation}
\frac{d\bar{u}}{d\tau}\approx\frac{\bar{a}_{0}\bar{u}}{2\nu}\left(1-\frac{\bar{u}^{2}}{\bar{a}_{0}}\right).
\end{equation}
This equation also has an exact solution
\begin{equation}
\bar{u}=\sqrt{\frac{\bar{a}_{0}}{2}}\exp\left(\frac{\bar{a}_{0}}{4\nu}\,\tau\right)\sqrt{{\rm sech}\left(\frac{\bar{a}_{0}}{2\nu}\,\tau\right)},
\end{equation}
which describes monotonic saturation of the ZF {[}Fig.~\ref{fig:4MT}(b),
green curve with squares{]}. This shows that, qualitatively, the ZF
saturation can be explained as a result of phase mixing that occurs
due to the primary-wave coupling to higher harmonics. For quantitative
predictions, a more rigorous approach is proposed below based on exploring
the drifton phase-space dynamics.

\subsection{Control parameter $\boldsymbol{N}$ in the GO limit}

\label{subsec:NonlinearMI_N}

\begin{figure*}
\includegraphics[width=1\textwidth]{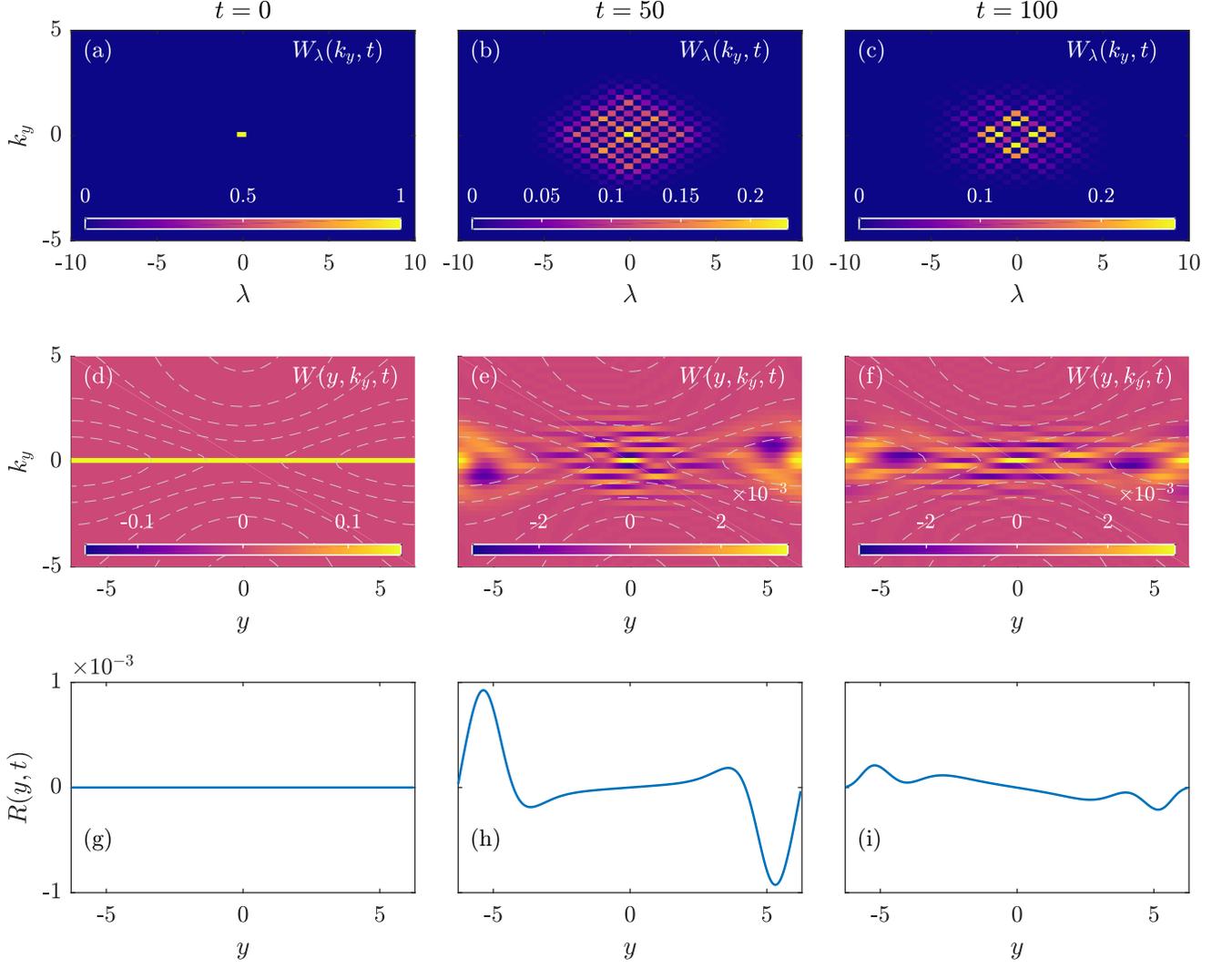}\caption{A numerical solution of Eq.~(\ref{eq:BasicEquation_Wigner_fullwave})
on the $(\lambda,k_{y})$ plane with constant $k_{x}=p$ and $257\times257$
grid points, namely, $-128\protect\leq m,n\protect\leq128$ {[}Eq.~(\ref{eq:NonlinearMI_4MT_deltas}){]}.
The initial condition is $\bar{W}_{0,0}=1$, and $\bar{W}_{m,n}=0$
for the rest. The ZF amplitude $u$ is kept constant. The parameters
are $\alpha=0,$ $\beta=1,$ $L_{{\rm D}}=1,$ $p=2,$ $q=0.5$, and
$u=0.05.$ (a)-(e): $|\bar{W}_{m,n}|$ at $t=0$, $50$, and $100$,
with $\lambda=mq$ and $k_{y}=nq/2$ at each grid point. (d)-(f):
the corresponding Wigner function $W(y,k_{y},t)$ {[}the inverse Fourier
transform of $W_{\lambda}$, Eq.~(\ref{eq:BasicEquation_Wigner_Spectral}){]}
at the same moments of time, where the gray dashed curves are constant-$\mathcal{H}$
contours {[}Eq.~(\ref{eq:BasicEquation_Wigner_hamiltonian}){]}.
The shape of the distribution function $W(y,k_{y})$ reflects the
presence of passing trajectories at $y=0$ and trapped trajectories
at $y=\pm\pi/q$. Since the initial condition is delta-shaped in $k_{y}$,
the GO assumption is not strictly satisfied, causing $W(y,k_{y},t)$
to be negative in some regions. (g)-(i): the ZF drive $R(y,t)$ {[}Eq.~(\ref{eq:NonlinearMI_MFlux}){]}
at the corresponding moments of time. It is seen that due to the passing
trajectories, the slope of $R$ at $y=0$ changes sign. (The associated
dataset is available at \protect\url{http://doi.org/10.5281/zenodo.2563449}.
\citep{zenodo}) \label{fig:gHME_regime1}}
\end{figure*}

\begin{figure}
\includegraphics[width=1\columnwidth]{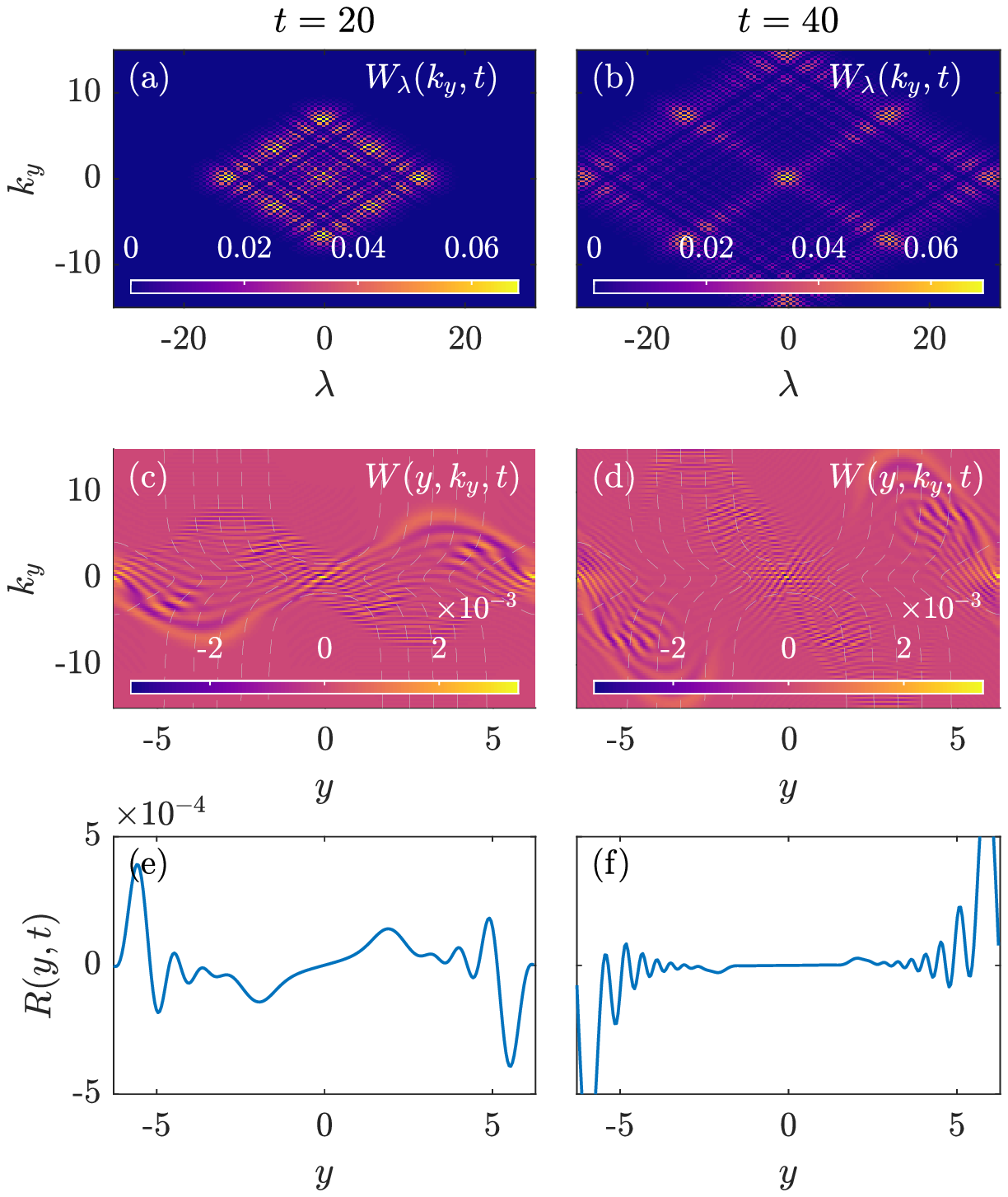}\caption{Same as Fig.~\ref{fig:gHME_regime1} but for $u=0.4$. First row:
$|\bar{W}_{m,n}|$ at $t=20$ and $t=40$. Second row: $W(y,k_{y},t)$
at the corresponding moments of time. The shape of the distribution
function $W(y,k_{y})$ reflects the presence of trapped trajectories
at $y=\pm\pi/q$ and runaway trajectories at larger-$|k_{y}|$ region.
Third row: the ZF drive $R(y,t)$ {[}Eq.~(\ref{eq:NonlinearMI_MFlux}){]}
at the corresponding moments of time. Due to runaways, the slope of
$R$ at $y=0$ quickly flattens and stays zero indefinitely. (The
associated dataset is available at \protect\url{http://doi.org/10.5281/zenodo.2563449}.
\citep{zenodo}) \label{fig:gHME_regime2}}
\end{figure}

Let us study the nonlinear stage of the MI using the WKE {[}Eq.~(\ref{eq:BasicEquation_WKE}){]},
which naturally accounts for all DW sidebands. As a starting point,
we appeal to the findings from Ref.~\citep{Zhu18b}, where three
types of drifton phase-space trajectories were identified: trapped,
passing, and runaway. (The last one corresponds to driftons leaving
to infinity along the $k_{y}$ axis while retaining finite $y$.)
It was also found in Ref.~\citep{Zhu18b} that the drifton phase-space
topology for a sinusoidal ZF can vary depending on how the ZF amplitude
relates to the following two critical values:
\begin{equation}
u_{c,1}\doteq\frac{\beta}{2p_{{\rm D}}^{2}-(\alpha L_{{\rm D}}^{-2}+q^{2})},\quad u_{c,2}\doteq\frac{\beta}{\alpha L_{{\rm D}}^{-2}+q^{2}}.\label{eq:NonlinearMI_WKE_uc}
\end{equation}
The GO limit corresponds to $u_{c,1}\ll u_{c,2}$. At $u<u_{c,1}$
(regime~1), trajectories of all three types are possible; at $u_{c,1}\leq u<u_{c,2}$
(regime~2), passing trajectories vanish; and finally, at $u>u_{c,2}$
(regime~3), trapped trajectories also vanish.

Figures~\ref{fig:gHME_regime1} and \ref{fig:gHME_regime2} show
the corresponding evolution of $W_{\lambda}$ calculated numerically
from Eq.~(\ref{eq:BasicEquation_Wigner_fullwave}) at fixed ZF velocity,
namely, $U=u\cos qy$ with constant $u$. Under the GO assumption,
the mHME and oHME exhibit similar dynamics, so for clarity, we consider
only the mHME ($\alpha=0$). Since $u$ is assumed stationary, Eq.~(\ref{eq:BasicEquation_Wigner_fullwave})
is linear in $W$, and the initial condition is chosen as {[}Eq.~(\ref{eq:NonlinearMI_4MT_deltas}){]}
\begin{equation}
\bar{W}_{m,n}(t=0)=\delta_{m,0}\delta_{n,0},
\end{equation}
where $\delta_{i,j}$ is the Kronecker delta. Figure~\ref{fig:gHME_regime1}
corresponds to $u<u_{c,1}$ (regime~1). In this case, the distribution
$W(y,\boldsymbol{k})$ in the phase space is confined to small $|k_{y}|$
regions enclosed by the contours of passing and trapped trajectories,
and no driftons reside on runaway trajectories (transport along $k_{y}$
is suppressed). The corresponding $\bar{W}_{m,n}$ are vanishingly
small at large $m$ and $n$; i.e., only a finite number of harmonics
are coupled. In contrast, Fig.~\ref{fig:gHME_regime2} corresponds
to $u_{c,1}<u<u_{c,2}$ (regime 2). In this case, passing trajectories
are replaced by runaway trajectories, so driftons can propagate to
much larger $|k_{y}|$ (transport along $k_{y}$ is \emph{not} suppressed).
Then, the corresponding $\bar{W}_{m,n}$ has a much wider distribution;
i.e., many harmonics are coupled simultaneously.

Now, let us consider the ZF evolution that would be driven by the
above dynamics (which we now consider prescribed for simplicity).
As seen from Eq.~(\ref{eq:BasicEquation_ZF}), the ZF is driven by
the gradient of the (minus) DW momentum flux,
\begin{equation}
R(y,t)\doteq-\langle\tilde{v}_{x}\tilde{v}_{y}\rangle(y,t),\label{eq:NonlinearMI_MFlux}
\end{equation}
which can be numerically calculated from the Wigner function \citep{Ruiz16},
\begin{equation}
R(y,t)=\int\frac{d^{2}kd\lambda}{(2\pi)^{3}}\,\frac{k_{x}k_{y}}{\kappa_{+\lambda}^{2}\kappa_{-\lambda}^{2}}\,W_{\lambda}(\boldsymbol{k},t)e^{i\lambda y}.\label{eq:NonlinearMI_Reynolds}
\end{equation}
The values of $R(y,t)$ are plotted in the lowest rows in Figs~\ref{fig:gHME_regime1}
and \ref{fig:gHME_regime2}. Since $\kappa_{\pm\lambda}^{2}=p_{{\rm D}}^{2}+(k_{y}\pm\lambda/2)^{2}$,
runaways with large $|k_{y}|$ contribute little to $R$, so the global
dynamics is largely determined by passing driftons. Particularly,
consider the slope of $R$ at the ZF peak ($y=0$). In regime~1,
this slope oscillates {[}Figs.~\ref{fig:gHME_regime1}(g)-(i){]},
hence causing oscillations of the ZF amplitude. In contrast, in regime
2, this slope quickly flattens and stays zero indefinitely {[}Figs.~\ref{fig:gHME_regime2}(g)-(i){]}.
This is due to the fact that driftons largely accumulate on runaway
and trapped trajectories near the ZF troughs ($y=\pm\pi/q$) and thus
cannot influence the ZF peak anymore.

Hence, whether a ZF will oscillate or saturate monotonically depends
on whether the ``control parameter'' $N\doteq u_{{\rm max}}/u_{c,1}$
is smaller or larger than unity. One can also make this estimate more
quantitative as follows. As discussed above, the DW spectrum is confined
to small $|k_{y}|$ when $N\lesssim1$; hence, the 4MT can be considered
as a reasonable model. By using Eq.~(\ref{eq:NonlinearMI_4MT_umax})
for $u_{{\rm max}}$, we obtain
\begin{equation}
u_{{\rm max}}\approx\frac{\sqrt{2}\gamma_{{\rm MI}}}{p},\quad u_{c,1}\approx\frac{\beta}{2p_{{\rm D}}^{2}},
\end{equation}
where we adopted $\delta,\delta'\gg1$ for the GO limit. Then, $N$
can be expressed as
\begin{equation}
N\approx\frac{2\sqrt{2}\,\gamma_{{\rm MI}}}{\omega_{{\rm DW}}},\quad\omega_{{\rm DW}}\doteq\frac{\beta p}{p_{{\rm D}}^{2}},\label{eq:NonlinearMI_N}
\end{equation}
where $\omega_{{\rm DW}}$ can be recognized as the (absolute value
of) the characteristic DW frequency. In summary, the ZF oscillates
if $\gamma_{{\rm MI}}\ll\omega_{{\rm DW}}$ and monotonically saturates
otherwise.

\subsection{Modifications due to full-wave effects}

\label{subsec:NonlinearMI_fullwave}

\begin{figure}
\includegraphics[width=1\columnwidth]{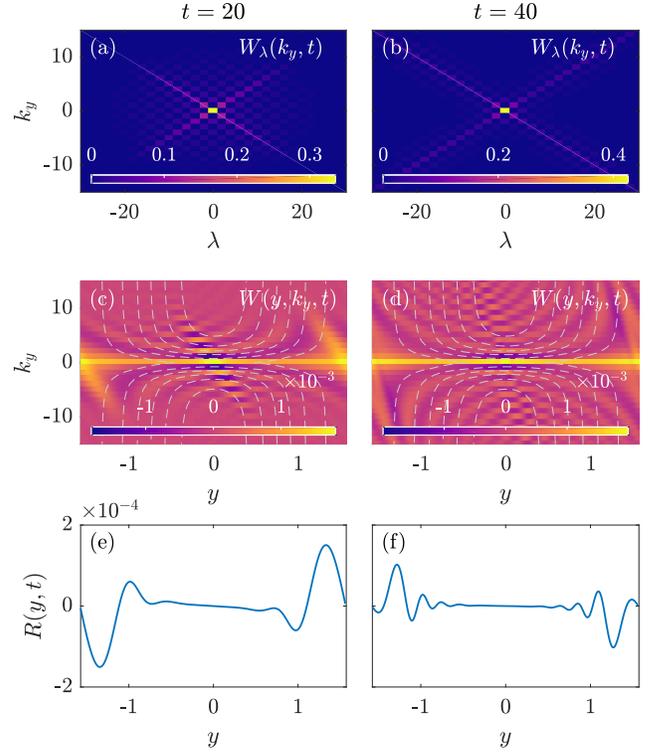}\caption{Same as Fig.~\ref{fig:gHME_regime1} but for $q=2$ and $u=0.2$.
Due to the large ZF wavenumber $q$, full-wave effects are significant.
Namely, $|\bar{W}_{m,n}|$ are localized along the diagonals $k_{y}=\pm\lambda/2$.
Also, $W(y,k_{y},t)$ cannot be easily interpreted as the drifton
distribution function. Instead, each drifton is smeared out in the
phase space. (The associated dataset is available at \protect\url{http://doi.org/10.5281/zenodo.2563449}.
\citep{zenodo}) \label{fig:gHME_full}}
\end{figure}

\begin{figure}
\includegraphics[width=1\columnwidth]{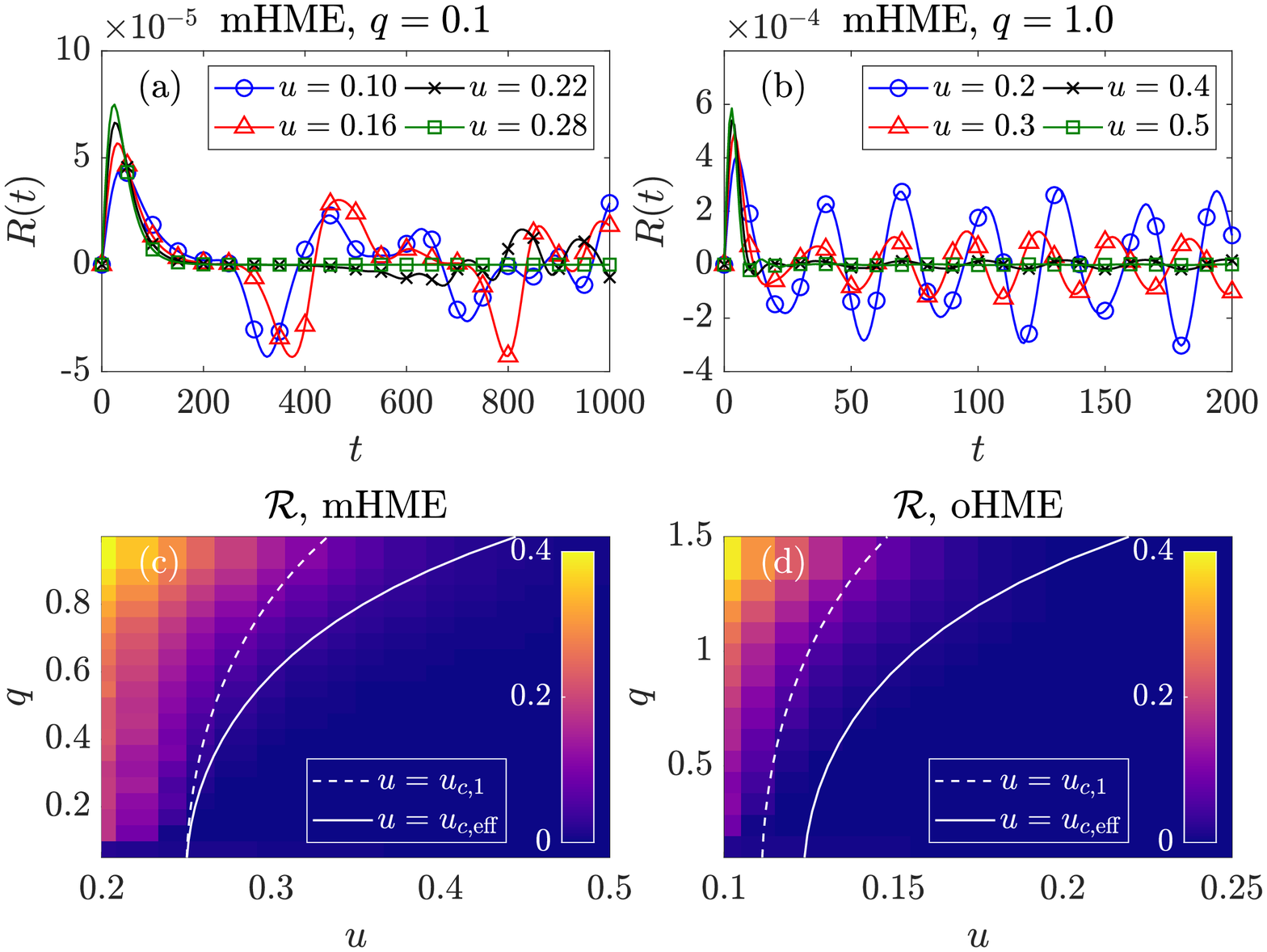}\caption{(a) and (b): Numerical solutions of Eq. (\ref{eq:BasicEquation_Wigner_fullwave})
at fixed ZF amplitude $u$ for $\alpha=0$ (mHME) and $\beta=L_{{\rm D}}=p=1$.
(a) The ZF drive $R(t)$ {[}Eq.~(\ref{eq:NonlinearMI_ZFDrive}){]}
at $y=0$ for various values of $u$ versus $t$ at $q=0.1$, which
corresponds to $u_{c,1}\approx0.25$ {[}Eq.~(\ref{eq:NonlinearMI_WKE_uc}){]}.
$R(t)$ oscillates when $u<u_{c,1}$ and decays to zero when $u>u_{c,1}$.
(b) Same as (a) but for $q=1.0$, which corresponds to $u_{c,1}\approx0.33$.
The critical amplitude, when $R(t)$ decays to zero, becomes larger
than $u_{c,1}$ and is estimated from Eq.~(\ref{eq:NonlinearMI_Full_ueff}).
(c) and (d): A parameter scan over $q$ and $u$ for determining the
critical ZF amplitude above which $R(t)$ decays to zero. The parameters
are $\beta=L_{{\rm D}}=1$, $\alpha=0$ (mHME) and $p=1$ in (c),
and $\alpha=1$ (oHME) and $p=2$ in (d). Shown in color is the value
of $\mathcal{R}$ {[}Eq.~(\ref{eq:NonlinearMI_Full_R}){]}. The white
dashed and solid curves are $u=u_{c,1}$ and $u=u_{c,{\rm eff}}$.
The latter roughly matches the threshold beyond which $\mathcal{R}$
is negligible. (The associated dataset is available at \protect\url{http://doi.org/10.5281/zenodo.2563449}.
\citep{zenodo}) \label{fig:Find_Uc}}
\end{figure}

At large enough $q$, the WKE (\ref{eq:BasicEquation_WKE}) ceases
to be valid and one must take into account the deviations from the
GO approximation. These deviations are called full-wave effects and
can be understood from the coupling between harmonics in Eq.~(\ref{eq:BasicEquation_Wigner_Spectral}).
At large $q$, the coupling coefficient $Q$ deviates from unity and
becomes inhomogeneous {[}Eq.~(\ref{eq:BasicEquation_Wigner_Q}){]}.
Recall that
\begin{equation}
\kappa_{a}^{2}\doteq p_{{\rm D}}^{2}+\left(k_{y}+\frac{a}{2}\right)^{2},
\end{equation}
hence the minimum value of $Q_{a}$ is achieved at $k_{y}+a/2=0$,
and 
\begin{equation}
\min Q=1-\delta_{\alpha}^{-1},\quad\delta_{\alpha}\doteq\frac{p_{{\rm D}}^{2}}{\alpha L_{{\rm D}}^{-2}+q^{2}}.\label{eq:NonlinearMI_delta_alpha}
\end{equation}
(Note that $\delta_{\alpha}>1$ is a necessary condition for the MI
to happen, and hence $\min Q>0$.) When $\min Q\ll1$, it can be seen
from Eq.~(\ref{eq:BasicEquation_Wigner_Q}) that the modes along
the diagonals $k_{y}=\pm\lambda/2$ are decoupled from the rest and
the initial perturbation $\bar{W}_{0,0}$ propagates mainly along
the diagonals, as demonstrated in Fig.~\ref{fig:gHME_full}.

When full-wave effects are important, the critical ZF amplitude can
deviate from $u_{c,1}$ given by Eq.~(\ref{eq:NonlinearMI_WKE_uc}).
Here, we study the critical ZF amplitude by numerically integrating
Eq.~(\ref{eq:BasicEquation_Wigner_fullwave}) with stationary $u$
and recording the ZF drive at $y=0$, namely,
\begin{equation}
R(t)\doteq\partial_{y}R(y,t)|_{y=0}=-\partial_{y}\langle\tilde{v}_{x}\tilde{v}_{y}\rangle(y,t)|_{y=0}.\label{eq:NonlinearMI_ZFDrive}
\end{equation}
Specifically, we calculate
\begin{equation}
\mathcal{R}\doteq\sqrt{\frac{\overline{R^{2}(t)}}{\max R^{2}(t)}},\label{eq:NonlinearMI_Full_R}
\end{equation}
where $\overline{\cdot\cdot\cdot}$ is the time-average. (To exclude
the initial transient dynamics, only the interval $0.25T<t<T$ is
used for the time-averaging, where $T$ is the total integration time.)
The critical amplitude is defined as the value of $u$ beyond which
$\mathcal{R}$ becomes negligible.

As shown in Figs.~\ref{fig:Find_Uc}(c) and (d), the critical amplitude
starts to deviate from $u_{c,1}$ as $q$ increases. The following
empirical correction can be adopted to account for the finite $q$-dependence
of the critical amplitude: 
\begin{equation}
u_{c,1}\to u_{c,{\rm eff}}\doteq\frac{u_{c,1}}{1-0.5\delta_{\alpha}^{-1}},\label{eq:NonlinearMI_Full_ueff}
\end{equation}
as seen in Figs.~\ref{fig:Find_Uc}(c) and (d). Then, the control
parameter (\ref{eq:NonlinearMI_N}) becomes
\begin{align}
N & =\frac{2\sqrt{2}\gamma_{{\rm MI}}}{\omega_{{\rm DW}}}\,\sqrt{\frac{\delta(\delta+1)}{\delta'(\delta'-1)}}\,\left(1-\frac{1}{2\delta_{\alpha}}\right)^{2},\label{eq:NonlinearMI_Full_N}
\end{align}
where $\delta,$ $\delta'$, and $\delta_{\alpha}$ are given by Eqs\@.~(\ref{eq:BasicEquation_delta_deltap})
and (\ref{eq:NonlinearMI_delta_alpha}). The ZF oscillates at $N\ll1$
and saturates at $N\gtrsim1$. 

\section{Numerical simulations}

\label{sec:Simulations}

\subsection{Parameter scan}

\label{subsec:Simulations_Scan}

\begin{figure}
\includegraphics[width=1\columnwidth]{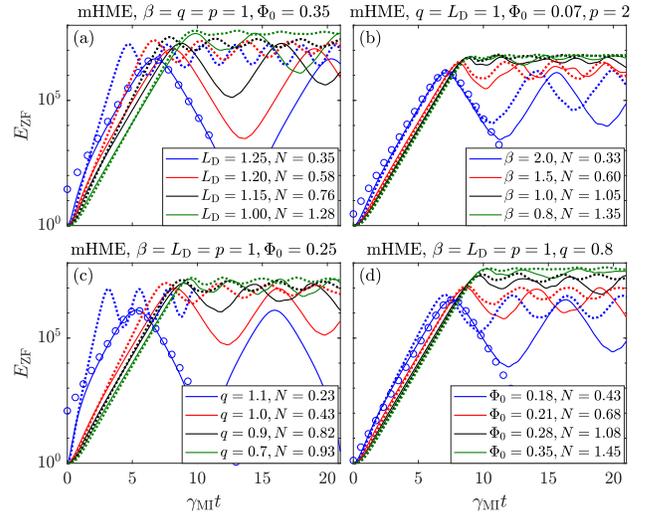}\caption{Numerical simulations of the QL (solid lines) and the NL mHWE (dotted
lines, using the same parameters). The blue circles are the corresponding
4MT solutions (\ref{eq:NonlinearMI_4MT_exact}). Shown is the ZF energy
$E_{{\rm ZF}}$ {[}Eq.~(\ref{eq:Simulations_GO_Ezf}){]} in units
$E_{{\rm ZF}}(t=0)$ versus time $t$ in units $\gamma_{{\rm MI}}^{-1}$
{[}Eq.~(\ref{eq:BasicEquation_gamma}){]}. The specific parameters
are presented in the corresponding figures. The initial conditions
are given by Eq.~(\ref{eq:Simulations_GO_initial}). The best agreement
between the QL and NL simulations can be found in the $\beta$-scan
{[}figure (b){]}, when the GO approximation is satisfied with the
highest accuracy ($\delta_{\alpha}=5$). (The associated dataset is
available at \protect\url{http://doi.org/10.5281/zenodo.2563449}.
\citep{zenodo}) \label{fig:Simulatsion_mHMEFigure}}
\end{figure}

\begin{figure}
\includegraphics[width=1\columnwidth]{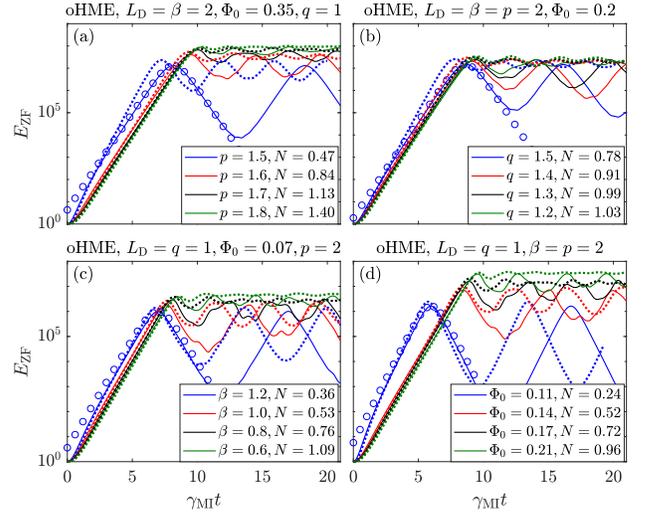}

\caption{Same as in Fig.~\ref{fig:Simulatsion_mHMEFigure}, but for the oHME
($\alpha=1$). Good agreement between the QL and NL simulations can
be found in the $\beta$-scan and $\Phi_{0}$-scan {[}figures (c)
and (d){]}, when the GO approximation is well satisfied ($\delta_{\alpha}=2.5$).
(The associated dataset is available at \protect\url{http://doi.org/10.5281/zenodo.2563449}.
\citep{zenodo}) \label{fig:Simulations_oHMEFigure}}
\end{figure}

\begin{figure}
\includegraphics[width=1\columnwidth]{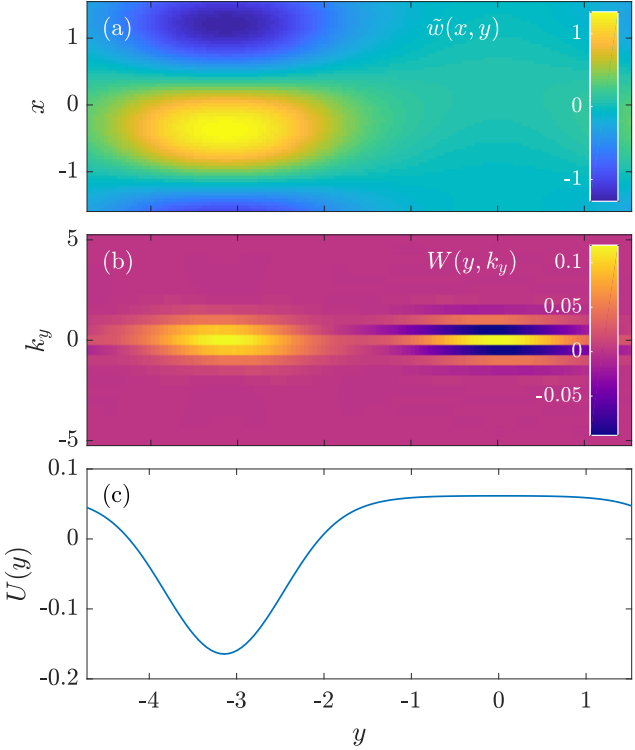}

\caption{A snapshot of the DW vorticity field $\tilde{w}(x,y)$, the corresponding
Wigner function $W(y,k_{y})$, and the ZF profile $U(y)$, at $\gamma_{{\rm MI}}t=7$
of the $\beta=2.0$ case in Fig.~\ref{fig:Simulatsion_mHMEFigure}(b)
(QL mHME simulations). At $U<0$, the vortex structure in $\tilde{w}$
corresponds to the trapped drifton distribution in $W$; at $U>0$,
the absence of $\tilde{w}$ corresponds to driftons following passing
trajectories and leaving this region. The striped structure in figure
(b) is due to the interference between the shown trapped distribution
and a similar distribution on the next spatial period (not shown).
Such structures are discussed in further detail in Ref.~\citep{soliton}.
(The associated dataset is available at \protect\url{http://doi.org/10.5281/zenodo.2563449}.
\citep{zenodo}) \label{fig:Simulations_Wignerfunction}}
\end{figure}

In order to test the above theory of the ZF fate beyond the linear
stage, we numerically integrated both the the QL system {[}Eqs.~(\ref{eq:BasicEquation_DW})
and (\ref{eq:BasicEquation_ZF}){]} and the NL system {[}Eqs.~(\ref{eq:BasicEquation_HME})
and (\ref{eq:BasicEquation_vorticity}){]} for various parameters
such that 
\begin{equation}
0<N\lesssim1,\quad\delta_{\alpha}\gtrsim1.\label{eq:Simulations_GO_params}
\end{equation}
The second requirement means $\delta_{\alpha}$ shall not be too large,
i.e., $q^{2}$ shall not be too small, because then ZF harmonics with
wave numbers that are multiples of $q$ have higher growth rates and
outpace the fundamental harmonic at the linear stage (see more discussions
in Sec.~\ref{subsec:Comparisons_NvsM}). Overall there are five parameters
that determine the system dynamics at the nonlinear stage: $L_{{\rm D}}$,
$p$, $\beta$, $q$, and $\Phi_{0}$. For the mHME ($\alpha=0$),
we vary only $L_{{\rm D}}$, $\beta$, $q$, and $\Phi_{0}$, because
$L_{{\rm D}}$ and $p$ mainly appear as a combination $L_{{\rm D}}^{-2}+p^{2}$,
and the remaining $p$ in Eq.~(\ref{eq:BasicEquation_Wigner_fullwave})
only defines the time scale and can be absorbed by a variable transformation
$t\to pt$. For the oHME ($\alpha=1$), we vary only $p$, $\beta$,
$q$, and $\Phi_{0}$, because it is hard to satisfy the requirement
(\ref{eq:Simulations_GO_params}) when $L_{{\rm D}}$ is varied with
other parameters fixed. For both the oHME and the mHME, the initial
conditions are chosen to be
\begin{equation}
\tilde{\varphi}=2\Phi_{0}\cos px,\quad U=u\cos qy,\quad u\ll p\Phi_{0},\label{eq:Simulations_GO_initial}
\end{equation}
where $\Phi_{0}$ is real

The simulation results of the QL and NL systems are shown in Figs.~\ref{fig:Simulatsion_mHMEFigure}
(mHME) and \ref{fig:Simulations_oHMEFigure} (oHME), where we plot
the ZF energy
\begin{equation}
E_{{\rm ZF}}(t)\doteq\frac{1}{2}\int dy[U(y,t)]^{2}\label{eq:Simulations_GO_Ezf}
\end{equation}
versus time $t$. As predicted by our theory, in QL simulations, ZFs
oscillate at the nonlinear stage if $N\ll1$ and largely saturate
monotonically if $N\gtrsim1$. For comparison, we also plot the corresponding
4MT solutions (\ref{eq:NonlinearMI_4MT_exact}) for the smallest-$N$
cases in each figure. The fact that the 4MT solutions agree with predictions
of the QL theory confirms the 4MT applicability at the nonlinear stage
of the MI in the weak-ZF limit.

Note that the ZF energy has a larger maximum at larger $N$. This
is explained as follows. At larger $N$, driftons propagate to larger
wave numbers and hence end up with lower energy, which is given by
\citep{Ruiz16}
\begin{equation}
E_{{\rm DW}}(t)\doteq\frac{1}{2}\int\frac{d^{2}kdy}{(2\pi)^{2}}\,\frac{W(y,\boldsymbol{k},t)}{k_{{\rm D}}^{2}}.\label{eq:Simulations_GO_Edw}
\end{equation}
Due to the total energy conservation, this leaves more energy for
the ZFs. 

In Figs.~\ref{fig:Simulatsion_mHMEFigure} and \ref{fig:Simulations_oHMEFigure},
NL simulation results are also shown for the same parameters. The
transition from ZF oscillations to saturation is also recovered from
these simulations. However, in terms of the oscillation amplitudes
and frequencies, good agreement between QL and NL simulations occurs
only when the GO approximation is satisfied with reasonable accuracy,
say, at
\begin{equation}
\delta_{\alpha}\gtrsim3.
\end{equation}
At smaller $\delta_{\alpha}$, NL simulations show much smaller amplitudes
of the ZF oscillations due to the additional DW\textendash DW self-interactions.
A brief explanation of the discrepancy between QL and NL simulations
is given in Sec.~\ref{subsec:Simulations_NL}.

In Fig.~\ref{fig:Simulations_Wignerfunction}, we show a snapshot
of the DW vorticity $\tilde{w}(x,y)$, the corresponding Wigner function
$W(y,k_{y})$, and the ZF profile $U(y)$ from a QL mHME simulation.
The snapshot is taken at $\gamma_{{\rm MI}}t=7$ of the $\beta=2.0$
case in Fig.~\ref{fig:Simulatsion_mHMEFigure}(b). This corresponds
to the time when the ZF energy reaches the maximum and is about to
reverse. The DW vortex structure at $U<0$ is clearly seen and corresponds
to trapped driftons in the phase space. In contrast, there is almost
no DW activity at $U>0$, because driftons follow passing trajectories
and have left this region. At this stage, the ZF velocity is no longer
sinusoidal, and have a deep trough and a flat peak. This shape of
the ZF is due to the larger ZF drive {[}Eq.~(\ref{eq:NonlinearMI_Reynolds}){]}
induced by the trapped trajectories at the ZF trough, and hence cause
the ZF to have a larger local amplitude. After this moment of time,
passing driftons return to the ZF top since the system is periodic
in $y$, and reduce the ZF amplitude; correspondingly, the ZF energy
oscillates, as in Fig.~\ref{fig:Simulatsion_mHMEFigure}(b).

\subsection{Difference between the QL and NL models}

\label{subsec:Simulations_NL}

Here, we discuss why good agreement between QL and NL systems can
be achieved when the GO approximation is well satisfied, and why discrepancies
arise otherwise. Recall that the MI growth rate $\gamma_{{\rm MI}}$
(\ref{eq:BasicEquation_gamma}) derived from the 4MT is exact for
the QL model but not for the corresponding NL model. Therefore, it
is expected that the discrepancy can be attributed, at least partly,
to the difference between the QL and NL growth rates. We compare these
growth rates by comparing the relative amplitude between the second
DW sideband and the first DW sideband at the linear stage, the former
being excluded from the 4MT. From Eq.~(\ref{eq:BasicEquation_4MT_Fourier}),
we have
\begin{equation}
\epsilon_{{\rm NL}}\doteq\left|\frac{\partial_{t}\varphi_{2\boldsymbol{p}+\boldsymbol{q}}}{\partial_{t}\varphi_{\boldsymbol{p}+\boldsymbol{q}}}\right|\sim\left|\frac{T(2\boldsymbol{p}+\boldsymbol{q},\boldsymbol{p}+\boldsymbol{q},\boldsymbol{p})\Phi_{0}\varphi_{\boldsymbol{p}+\boldsymbol{q}}}{T(\boldsymbol{p}+\boldsymbol{q},\boldsymbol{q},\boldsymbol{p})\Phi_{0}\varphi_{\boldsymbol{q}}}\right|
\end{equation}
as a measure of the importance of NL effects. Also, $|\varphi_{\boldsymbol{p}}|=|\Phi_{0}|$
is assumed at the linear stage. From Eq.~(\ref{eq:BasicEquation_4MT_coefficient}),
the coupling coefficients are
\begin{gather*}
|T(2\boldsymbol{p}+\boldsymbol{q},\boldsymbol{p}+\boldsymbol{q},\boldsymbol{p})|=\frac{q^{2}}{L_{{\rm D}}^{-2}+4p^{2}+q^{2}}\,pq,\\
|T(\boldsymbol{p}+\boldsymbol{q},\boldsymbol{q},\boldsymbol{p})|=\frac{(1-\alpha)L_{{\rm D}}^{-2}+p^{2}-q^{2}}{L_{{\rm D}}^{-2}+p^{2}+q^{2}}\,pq.
\end{gather*}
Then
\begin{equation}
\epsilon_{{\rm NL}}\sim\left|\frac{\varphi_{\boldsymbol{p}+\boldsymbol{q}}}{\varphi_{\boldsymbol{q}}}\right|\frac{q^{2}}{L_{{\rm D}}^{-2}+4p^{2}+q^{2}}\,\frac{L_{{\rm D}}^{-2}+p^{2}+q^{2}}{(1-\alpha)L_{{\rm D}}^{-2}+p^{2}-q^{2}}.
\end{equation}
In the above expression for $\epsilon_{{\rm NL}}$, the first term
$|\varphi_{\boldsymbol{p}+\boldsymbol{q}}/\varphi_{\boldsymbol{q}}|$
can be estimated from the 4MT equation (\ref{eq:BasicEquation_4MTeq}),
which gives
\begin{equation}
\left|\frac{\partial_{t}\varphi_{\boldsymbol{p}+\boldsymbol{q}}}{\partial_{t}\varphi_{\boldsymbol{q}}}\right|\sim\left|\frac{T(\boldsymbol{p}+\boldsymbol{q},\boldsymbol{q},\boldsymbol{p})\Phi_{0}\varphi_{\boldsymbol{q}}}{T(\boldsymbol{q},-\boldsymbol{p},\boldsymbol{p}+\boldsymbol{q})\Phi_{0}\varphi_{\boldsymbol{p}+\boldsymbol{q}}}\right|.
\end{equation}
Since $\partial_{t}\varphi_{\boldsymbol{p}+\boldsymbol{q}}/\partial_{t}\varphi_{\boldsymbol{q}}\sim\varphi_{\boldsymbol{p}+\boldsymbol{q}}/\varphi_{\boldsymbol{q}}$
and $|T(\boldsymbol{q},-\boldsymbol{p},\boldsymbol{p}+\boldsymbol{q})|=pq$,
this gives
\begin{equation}
\left|\frac{\varphi_{\boldsymbol{p}+\boldsymbol{q}}}{\varphi_{\boldsymbol{q}}}\right|\sim\sqrt{\frac{(1-\alpha)L_{{\rm D}}^{-2}+p^{2}-q^{2}}{L_{{\rm D}}^{-2}+p^{2}+q^{2}}},
\end{equation}
and thus
\begin{equation}
\epsilon_{{\rm NL}}\sim\frac{q^{2}}{L_{{\rm D}}^{-2}+4p^{2}+q^{2}}\left(\frac{\delta+1}{\delta_{\alpha}-1}\right)^{1/2},
\end{equation}
where $\delta$ and $\delta_{\alpha}$ are given by Eqs\@.~(\ref{eq:BasicEquation_delta_deltap})
and (\ref{eq:NonlinearMI_delta_alpha}), respectively. When $\delta,\delta_{\alpha}\gg1$,
i.e., $q^{2}\ll p^{2}+(1-\alpha)L_{{\rm D}}^{-2}$, one obtains $\epsilon_{{\rm NL}}\ll1$.
Then, NL effect is expected to be small, and hence QL and NL simulations
produce similar results. In contrast, when $\delta_{\alpha}$ approaches
unity, $\epsilon_{{\rm NL}}$ can become of order one. Then, the QL
model ceases to be an adequate approximation to the NL model.

The above estimate also indicates that, at least for the mHME, when
$p^{2}\gtrsim L_{{\rm D}}^{-2}$, one has
\begin{equation}
\epsilon_{{\rm NL}}\sim\frac{\epsilon_{{\rm GO}}}{4}\sqrt{\frac{1+\epsilon_{{\rm GO}}}{1-\epsilon_{{\rm GO}}}},
\end{equation}
where $\epsilon_{{\rm GO}}\sim\delta^{-1}\ll1$ under the GO approximation.
Therefore, $\epsilon_{{\rm NL}}\ll1$ at $\epsilon_{{\rm GO}}\ll1$
and $\epsilon_{{\rm NL}}\gtrsim1$ when $\epsilon_{{\rm GO}}$ approaches
unity (and the MI vanishes when $\epsilon_{{\rm GO}}>1$). Hence,
the applicability domain of the QL approximation is roughly the same
as that of the GO approximation.

\section{Comparison with previous studies}

\label{sec:Comparisons}

\begin{figure}
\includegraphics[width=1\columnwidth]{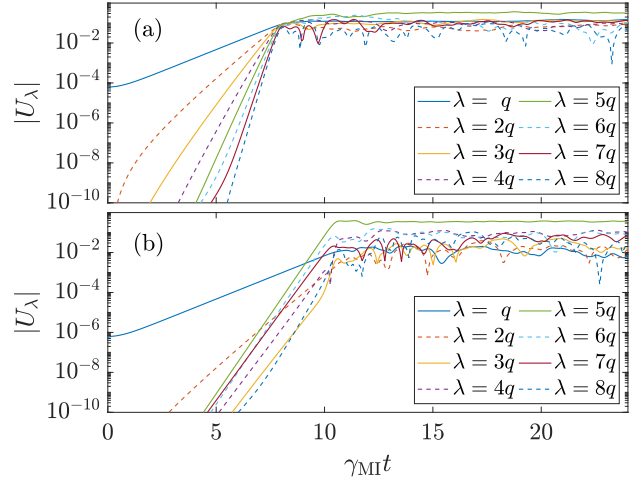}\caption{Numerical simulations of the QL oHME, which corresponds to the $M=1$
case in Ref.~\citep{Connaughton10}. The parameters are $L_{{\rm D}}^{-2}=0,$
$\beta=p=10$, $q=1$, and $\Phi_{0}=10^{-2}$. The initial ZF perturbation
is $U(y,t=0)=u\cos qy$. Shown are the amplitudes of the ZF harmonics
$|U_{\lambda}|$ for different $\lambda$: (a) $u=2\times10^{-5}$
and (b) $u=2\times10^{-7}$. The time $t$ is normalized to $\gamma_{{\rm MI}}^{-1}$
of the fundamental harmonic $\lambda=q$. Since higher harmonics have
larger MI growth rates, they grow faster and eventually, in (b), outpace
the fundamental harmonic. For the larger initial ZF perturbation in
(a), which is also assumed in Ref.~\citep{Connaughton10}, higher
harmonics are not as strong as in (b) because the linear stage ends
sooner. Nevertheless, higher harmonics around $\lambda=5q$ still
dominate at the nonlinear stage. Therefore, one should replace $q$
by $5q$ in the calculation of $N$ in order to achieve the correct
prediction. (The associated dataset is available at \protect\url{http://doi.org/10.5281/zenodo.2563449}.
\citep{zenodo}) \label{fig:Comparison_M=00003D1}}
\end{figure}

\subsection{Control parameters: $\boldsymbol{N}$ versus $\boldsymbol{M}$}

\label{subsec:Comparisons_NvsM}

Let us also compare our results with those in Refs.~\citep{Connaughton10,Gallagher12}
where related simulations were performed. Specifically, Ref. \citep{Connaughton10}
reports NL oHME simulations for $p^{2}\gg q^{2}$ and $L_{{\rm D}}^{-2}=0$.
Also, Ref. \citep{Gallagher12} reports NL mHME simulations; however,
the parameter is chosen such that $p^{2}\gg q^{2},L_{{\rm D}}^{-2}$,
and the resulting dynamics is almost identical to that in the oHME.
In both cases, the GO assumption is well satisfied ($\delta,\delta',\delta_{\alpha}\approx100\gg1$).
Hence, the difference between NL and QL simulations is expected to
be small (see Sec.~\ref{subsec:Simulations_NL}), and the results
of Refs.~\citep{Connaughton10,Gallagher12} can be compared with
ours within the scope of the QL approximation. (We have indeed been
able to reproduce all the related results from QL simulations using
the same parameters as in Refs.~\citep{Connaughton10,Gallagher12},
but we choose not to duplicate the figures here.) Due to the similar
choice of the parameters in Ref.~\citep{Connaughton10} and Ref.~\citep{Gallagher12},
we compare with Ref.~\citep{Connaughton10} only.

Within the GO limit, our control parameter $N$ {[}Eq. (\ref{eq:NonlinearMI_Full_N}){]}
is
\begin{equation}
N\approx\frac{2\sqrt{2}\,p\gamma_{{\rm MI}}}{\beta},\quad\gamma_{{\rm MI}}\approx\sqrt{2p^{2}q^{2}|\Phi_{0}|^{2}-\left(\frac{\beta q^{2}}{p^{3}}\right)^{2}}.\label{eq:Comparison_N}
\end{equation}
A different parameter was proposed in Ref.~\citep{Connaughton10},
namely, 
\begin{equation}
M\doteq\frac{p^{3}\Phi_{0}}{\beta}.\label{eq:Comparison_M}
\end{equation}
The authors argue that $M\ll1/3$ corresponds to ZF oscillations and
$M\gtrsim1/3$ corresponds to monotonic ZF saturation. However, in
contrast to our quantitative derivations, only a qualitative argument
is provided in Ref.~\citep{Connaughton10} (also see Sec.~\ref{subsec:Comparisons_RK}).
As a result, the parameter $M$ cannot describe the sensitive dependence
on $q$ or $L_{{\rm D}}$, as our $N$ does in Figs.~\ref{fig:Simulatsion_mHMEFigure}(a),
\ref{fig:Simulatsion_mHMEFigure}(c), and \ref{fig:Simulations_oHMEFigure}(b).
Our $N$ is also better in terms of the predictive power. For example,
for the case of $\beta=1.2$ in Fig.~\ref{fig:Simulations_oHMEFigure}(c),
one has $M=0.47>1/3$, so the ZF is supposed to saturate; however,
the ZF actually oscillates, which can be predicted by our $N=0.36<1$.

We note that our theory can also predict the outcomes of three numerical
examples used in Ref.~\citep{Connaughton10}, where $M=0.1$, $1$,
and $10$, respectively. For $M=0.1$ and $M=10$, our control parameter
takes the values $N=0.028$ and $N=3.94$, respectively; hence, our
theory predicts that the ZF oscillates in the former case and saturates
monotonically in the latter case, which is indeed observed in Ref.~\citep{Connaughton10}.
However, Ref.~\citep{Connaughton10} reports ZF saturation at $M=1$,
while our $N=0.39<1$. It may seem then that our theory predicts ZF
oscillations instead of saturation, which would be incorrect. Actually,
due to the small $q$ in this case, high harmonics of the ZF have
much larger linear growth rates and outpace the fundamental harmonic
(Fig.~\ref{fig:Comparison_M=00003D1}). This makes our analytic model
inapplicable, for it assumes a quasi-sinusoidal ZF. That said, if
one calculates $N$ by replacing $q$ with the wave number of the
dominant harmonic, then $N$ becomes larger than unity and the ZF
saturation is readily anticipated.

\subsection{Relevance of the Rayleigh\textendash Kuo threshold for the ZF saturation}

\label{subsec:Comparisons_RK}

The authors of Ref.~\citep{Connaughton10} proposed a brief explanation
of the physical meaning of their parameter $M$, which is as follows.
First, they assumed that DWs transfer an order-one fraction of their
energy to ZFs, so the ZF maximum amplitude is
\begin{equation}
u_{{\rm max}}\sim p\Phi_{0}.\label{eq:Compairson_RK_umax}
\end{equation}
Second, they speculated that the ZF saturates when it reaches the
Rayleigh\textendash Kuo (RK) threshold \citep{Kuo49}
\begin{equation}
\partial_{y}^{2}U(y)-\beta>0.
\end{equation}
We believe that this explanation is problematic, namely, for two reasons.
First, the estimate (\ref{eq:Compairson_RK_umax}) contradicts the
4MT estimate (\ref{eq:NonlinearMI_4MT_umax}) that we have confirmed
numerically (Figs.~\ref{fig:Simulatsion_mHMEFigure} and \ref{fig:Simulations_oHMEFigure}).
A more accurate estimate within the GO regime would be $u_{{\rm max}}\gtrsim2q\Phi_{0}$.
(Note that this estimate reinstates the dependence on $q$, which
is absent in $M$.) Second, the RK criterion does not describe the
ZF saturation but rather determines the threshold of the instability
of the Kelvin\textendash Helmholtz type that destroys the ZF. (It
is also called the ``tertiary instability'' by some authors \citep{Rogers00,Zhu18a,Zhu18b,Kuo49,Zhu18c,Numata07,Kim2002,Rath18},
and in previous studies we showed that this instability does not exist
in the GO limit \citep{Zhu18a,Zhu18b,Zhu18c}.) Since the RK threshold
corresponds to $u>u_{c,2}$ (Sec.~\ref{subsec:NonlinearMI_N}), and
$u_{c,2}\gg u_{c,1}$ in the GO regime, we claim that ZFs saturate
before the RK threshold is reached. In summary, we believe that our
parameter $N$ is more substantiated than the parameter $M$ introduced
in Refs.~\citep{Connaughton10,Gallagher12}. We also emphasize that
our theory withstands the test of numerical simulations, as shown
in Sec.~\ref{subsec:Simulations_Scan}.

\section{Conclusions}

\label{sec:conclusion}

In this paper we propose a semi-analytic theory that explains the
transition between the oscillations and saturation of collisionless
ZFs within the QL HME. By analyzing phase-space trajectories of driftons
within the GO approximation, we argue that the parameter that controls
this transition is $N\sim\gamma_{{\rm MI}}/\omega_{{\rm DW}}$, where
$\gamma_{{\rm MI}}$ is the MI growth rate and $\omega_{{\rm DW}}$
is the linear DW frequency. We argue that at $N\ll1$, ZFs oscillate
due to the presence of so-called passing drifton trajectories, and
we derive an approximate formula for the ZF amplitude as a function
of time in this regime. In doing so, we also extend the applicability
of the popular 4MT model, which is commonly used for the linear stage,
to nonlinear ZF\textendash DW interactions. We also show that at $N\gtrsim1$,
the passing trajectories vanish and ZFs saturate monotonically, which
can be attributed to phase mixing of higher-order sidebands. A modification
of $N$ that accounts for effects beyond the GO limit is also proposed.
These analytic results are tested against both QL and NL simulations.
They also explain the earlier numerical results by Refs.~\citep{Connaughton10,Gallagher12}
and offer a revised perspective on what the control parameter is that
determines the transition from oscillations to saturation of collisionless
ZFs.
\begin{acknowledgments}
This work was supported by the U.S. Department of Energy (DOE), Office
of Science, Office of Basic Energy Sciences, and also by the U.S.
DOE through Contract No. DE-AC02-09CH11466.
\end{acknowledgments}

\end{document}